\documentclass{aastex631}

\usepackage{placeins}
\usepackage{hyperref}
\begin{document}
\shorttitle{TESS Blazar Light Curves}
\shortauthors{Poore et al.}

\title{A Comparative Study of TESS Light Curve Extraction Methods Applied to Blazars}

\author[0009-0001-1226-8640]{Ethan Poore}
\affiliation{Western Kentucky University,
Department of Physics and Astronomy,
1906 College Heights Blvd.,
Bowling Green, KY 42101, USA}
\affiliation{Texas A\&M University,
576 University Dr,
College Station, TX 77840, USA}

\author[0000-0001-8961-2465]{Michael Carini}
\affiliation{Western Kentucky University,
Department of Physics and Astronomy,
1906 College Heights Blvd.,
Bowling Green, KY 42101, USA}

\author[0009-0007-0454-6245]{Ryne Dingler}
\affiliation{Texas A\&M University,
576 University Dr,
College Station, TX 77840, USA}
\affiliation{Southern Methodist University,
3215 Daniel Ave,
Dallas, TX 75205, USA}

\author[0000-0003-4737-1477]{Ann E.~Wehrle}
\affiliation{Space Science Institute, 4765 Walnut Street, Suite B,
Boulder, CO 80301, USA}

\author[0000-0002-1029-3746]{Paul J.~Wiita}
\affiliation{The College of New Jersey,  
Department of Physics, 
2000 Pennington Rd., 
Ewing, NJ 08628-0718, USA}

\begin{abstract}

{Blazars are characterized by largely aperiodic variability on timescales ranging from minutes to decades across the electromagnetic spectrum. The TESS (Transiting Exoplanet Survey Satellite) mission provides continuous sampling of blazar variability on timescales ranging from tens of minutes to 27 days for a single sector observation.  Proper removal of the background, thermal ramping, and onboard systematic effects are crucial to the extraction of a reliable blazar light curve. Multiple publicly available procedures have been created to correct for these effects. Using ground based observations from the Zwicky Transit Facility (ZTF) and the Asteroid Terrestrial-impact Last Alert System (ATLAS) as ``ground truth'' observations, we compare 6 different methods (Regression, Cotrending Basis Vectors (CBV), Pixel Level Decorrelation (PLD), \texttt{eleanor}, \texttt{quaver}, and simple differential photometry (SDP)) to each other, and to our ``ground truth'' observations, to identify which methods properly correct light curves of a sample of 11 bright blazars, including the suspected neutrino source TXS~0506+056. In addition to comparing the resulting light curves, we compare the slopes of the power spectral densities, perform least-square fitting to simultaneous ZTF data, and quantify other statistical qualities generated from the light curves of each method. We find that only three of the six methods compared (Simple Differential Photometry, \texttt{eleanor}, and  \texttt{quaver}) produce TESS light curves consistent with the ground-based ZTF and ATLAS observations.}

\end{abstract}



\setcounter{section}{0}

\section{Introduction}
    \label{sec:introduction}

Blazars are the most extreme example of the Active Galactic Nuclei (AGN) phenomenon. They display continuum variability across the electromagnetic spectrum on timescales from minutes to decades. The \textit{Kepler}, K2, and now the Transiting Exoplanet Survey Satellite (TESS) missions have greatly improved our understanding of the most rapid observed optical variability in these sources by providing high cadence, long-duration observations of blazars at optical wavelengths \citep[and Dingler et al. (Submitted to ApJ)]{2013ApJ...766...16E,2019ApJ...877..151W, 2020ApJ...900..137W, 2021MNRAS.501.1100R,2021MNRAS.504.5629R, 2023MNRAS.518.1459P}. TESS \citep{2015JATIS...1a4003R} continuously monitors a $24\degr \times 90\degr$ sector of the sky in the wavelength band of 6000\AA -- 10000\AA, shifting to different sectors after approximately 28 days. During the first two years of the TESS mission a Full-Frame Image (FFI) of the entire field of view (FOV) was obtained every 30 minutes, with postcard cutouts of observer-selected targets obtained at a higher cadence of 2 minutes. Beyond systematic trends, one must also correct TESS light curves for earth and moon background light, depending on the sector. These backgrounds can vary from 2--3 times the usual background of $\sim 100$ cts s$^{-1}$ to saturation as the TESS spacecraft moves through its earth-centered orbit twice every 28 days.

Extraction of light curves showing rapid optical variability, as seen in blazars, was not optimized in the TESS mission pipeline reduction software, as blazar science was ``bonus science'' from the TESS mission. TESS has no absolute photometric calibration. The TESS mission provides light curves, with source brightness in counts$/$second, for the high cadence (2-minute) observations in the form of simple aperture photometry (SAP) and a pre-search data conditioning pipeline (PDCSAP). PDCSAP corrects the SAP light curves for systematic effects, removes isolated outliers, corrects the flux of a source for crowding effects (the pixel size is 21\arcsec), and corrects for the standard aperture not containing all of the flux from a target source. However, the PDC process is known to remove longer-term variations common in blazar light curves. Extraction of light curves from the FFI images is left to the user. \citet{2020ApJ...900..137W}, \citet{2021MNRAS.501.1100R,2021MNRAS.504.5629R}, and \citet{2023MNRAS.518.1459P} have used TESS blazar PDCSAP light curves, while \citet{2023ApJ...943...53K} employed SAP light curves, and noted good agreement between ground-based observations of BL Lac and a PDCSAP-produced light curve. \citet{2021MNRAS.501.1100R} noted very little difference between SAP, PDCSAP, and ground-based light curves for the source S5 0716+714 but did note differences between the SAP and PDCSAP light curves for the source S4 0954+658, which was three magnitudes fainter than S5 0716+714 \citep{2021MNRAS.504.5629R}. Only 29 out of the nearly 100 of the blazars readily detectable with TESS have been targets of 2-minute cadence observations and thus have SAP and PDCSAP light curves available. The astronomical community has expended considerable effort to create methods to extract non-periodic light curves from both K2 \citep{2016AJ....152..100L, 2018AJ....156...99L,2014PASP..126..948V,2018ApJ...857..141S} and TESS \citep{2019PASP..131i4502F,2021tsc2.confE.188R, 2021ApJ...908...51F, 2023arXiv231008631S, 2023AJ....165...71H}. Dingler et al. (Submitted to ApJ) used the \texttt{quaver} \citep{2023arXiv231008631S} pipeline to examine a sample of 67 blazars observed with TESS. They were able to extract light curves which demonstrated ample variability from 31 objects from their sample under the ``simple" hybrid method of \texttt{quaver} and an additional two objects using the ``full" hybrid method. It has been shown by Dingler et al. (Submitted to ApJ), that the ``simple" extraction method shows better agreement with ground-based observation; however, it may not fully remove contamination from systematics or backgrounds. Meanwhile, the ``full" hybrid method is more capable of removing systematic and background trends, but may remove some likely-real variability and may need re-scaling to match with ground-based observations.

In this paper, we report the results of a comparative study of six methods of producing blazar light curves from TESS observations for a sample of 11 bright blazars, including the suspected neutrino source, TXS~0506+056. We examine methods available as part of the \textit{corrector class} in the \texttt{lightkurve} package\footnote{https://heasarc.gsfc.nasa.gov/docs/tess/NoiseRemovalv2.html}, which are methods adapted from the \textit{Kepler}/K2 mission (Regression, Cotrending Basis Vectors (CBV) and Pixel Level Decorrelation (PLD)), along with \texttt{eleanor} \citep{2019PASP..131i4502F},  \texttt{quaver} \citep{2023arXiv231008631S}, and simple differential photometry (SDP). The last three of those methods were chosen to be included in this comparison because they are currently the most typically used methods to extract blazar light curves from TESS observations. In addition to these six methods, \texttt{ISIS} \citep{2021ApJ...908...51F} and \texttt{TESSReduce} \citep{2021tsc2.confE.188R} provide the extraction of TESS light curves utilizing an image subtraction approach, which is markedly different from the approach taken by the methods we have chosen to compare. Image subtraction allows for the correction of systematics at the pixel level,and is optimized for studying supernovae.  For this paper, focused on blazars, we compared the light curves from the six extraction methods to each other and to ground truth light curves from the Zwicky Transient Facility (ZTF)  \citep{2019PASP..131a8002B} and the Asteroid Terrestrial-Impact Last Alert System (ATLAS) project \citep{2018PASP..130f4505T}.

In Section \ref{sec:targets}, we describe the target selection for this study. Section \ref{sec:extraction} discusses the six different light curve extraction methods. In Section \ref{sec:gb_obs}, we describe the ground truth observations used in this study and present the light curves from each extraction method, along with the ground truth observations. Section \ref{sec:results} gives the results of our comparison of the different methods, and Section \ref{sec:conclusions} summarizes our main conclusions. 
\FloatBarrier
\section{Target selection}
    \label{sec:targets}

We chose these eleven blazars because they were bright, likely variable, relatively free of nearby stars,  and, in the case of 3C 371, well studied over many years.  We chose blazars that were bright enough (TESS Input Catalog (TIC) magnitudes brighter than 15.5) to have good ground-based ZTF and/or ATLAS data even if they had faded significantly since the observations in various catalogs that were used in compiling the TIC.

We note that  simple differential photometry, quaver, and eleanor produced good results for the five blazars  with low count rates (30-60 cts~s$^{-1}$).  We do not know how our conclusions would change for much fainter sources because TESS is optimized for stars brighter than 15th magnitude:  instrumental effects plus the variable background ({$\sim$}100 cts~s$^{-1}$) may not be straightforward to extrapolate to fainter magnitudes. For stars, the  Combined Differential Photometric Precision (CDPP) as a function of TESS magnitude from 2 to 16 varies by sector, as shown in each sector’s Data Release Notes.  See, for example, the (CDPP) figure for Sector 1 in \url{ https://heasarc.gsfc.nasa.gov/docs/tess/observing-technical.html} or the individual Data Release Notes on MAST \url{https://archive.stsci.edu/tess/tess_drn.html. } A 16.5 magnitude target should typically yield approximately 43 cts s$^{-1}$pixel$^{-1}$ with an average background of approximately 13 cts s$^{-1}$pixel$^{-1}$  over the course of a 28-day sector observation. The backgrounds observed by each of the four cameras that make up the FOV are not identical. The list of targets used in this study is presented in Table \ref{tab:TESS_targets}. We give the target common name, TIC ID, redshift, blazar class, the TESS observing cycle, and the sector number. All of our selected sources are members of the BL Lac class of blazars. We note that \citet{2019MNRAS.484L.104P} have proposed that the suspected neutrino source TXS~0506+056 is a FSRQ masquerading as a BL Lac whose broad emission lines are so weak they are overwhelmed by the synchrotron emission of the jet. (BL Lacs and FSRQs have emission line equivalent widths less than and greater than 5\AA, respectively.)  Table \ref{tab:TESS_dates} presents observation details of the sectors listed in Table \ref{tab:TESS_targets}, including the calendar dates of the sectors, the corresponding MJD date range and the total number of observing days in the sector, excluding gap in the middle of the observations which occurs due to a pause in observations for data download.
\begin{deluxetable}{cccccccc}
        \label{tab:TESS_targets}

\tablecaption{Targets
}
\tablewidth{0pt}
\tablecolumns{10}
\tabletypesize{\small}
\tablehead{
         \colhead{Name}
         &\colhead{TIC ID}
         & \colhead{TESS Mag}
         & \colhead{$z$}
         & \colhead{Class}
         & \colhead{Cycle}
         & \colhead{Sector}
        \\
         \colhead{}
         &\colhead{}
         & \colhead{}
         & \colhead{}
         & \colhead{}
         & \colhead{}
         & \colhead{}
         }
\startdata
PKS~0048-097	 & 	3964982	 & 	15.2203	 & 	0.633 & BL Lac & 1 & 3	\\
PKS~0301-243	 & 	88416580	 & 	15.468	 & 	0.265	 & 	BL Lac & 1 & 4\\
PKS~0422+004	 & 	448808952	 & 	15.1783	 & 	0.310	 & 	BL Lac & 1 & 5\\
TXS~0506+056	 & 	455069576	 & 	14.7609	 & 0.3365 & BL Lac & 1 & 5 	\\
1ES~1011+496	 & 	156407432	 & 	15.1783	 & 	\nodata	 & 	BL Lac & 2 & 21	\\
ON~325	 & 	139069680	 & 	14.148	 & 0.13	 & BL Lac & 2 & 22		\\
ON~246	 & 	1153022	 & 	15.0705	 & 	0.135 & BL Lac & 2 & 22\\
PG~1246+586	 & 	150343823	 & 15.0467	 & 	0.84739	 & 	BL Lac & 2 & 21	\\
OQ~530	 & 	445860589	 & 	14.924	 & 	0.153	 & 	BL Lac & 2 & 15		\\
PKS~1424+240	 & 	424782650	 & 	13.5956	 & 	0.16	 & 	BL Lac & 2 & 23	\\
3C~371	 & 	219113011	 & 	14.4042	 & 	\nodata & BL Lac & 1 & 14 \\
\\
\enddata
\end{deluxetable}

\begin{deluxetable}{cccr}
    \label{tab:TESS_dates}
\tablecaption{Dates of TESS Observations}
\tablewidth{0pt}
\singlespace
\tablecolumns{4}
\tabletypesize{\small}
\tablehead{
         \colhead{TESS Sector}
         & \colhead{Dates}
         & \colhead{MJD}
         &\colhead{Science Days} 
         \\
         }
\startdata
3 & 2018 Sept 20 -- 2018 Oct 17 & 58381-58408 & 20.40 days \\
4 & 2018 Oct 19 -- 2018 Nov 14 & 58410-58435 & 22.23 days \\
5 & 2018 Nov 15 -- 2018 Dec 11 & 58438-58463 & 25.23 days \\
14 & 2019 Jul 18 -- 2019 Aug 14 & 58682-58709 & 25.91 days \\
15 & 2019 Aug 15 -- 2019 Sept 10 & 58710-58736 & 24.97 days \\
21 & 2020 Jan 21 -- 2020 Feb 18 & 58869-58897 & 24.42 days \\
22 & 2020 Feb 19 -- 2020 Mar 17 & 58900-58925 & 26.13 days \\
23 & 2020 Mar 19 -- 2020 Apr 15 & 58930-58954 & 25.81 days \\
\enddata
\end{deluxetable}
\FloatBarrier
\section{Light Curve Extraction Methods}
    \label{sec:extraction}

This section briefly describes the six different data reductions and the procedures used consistently during the study. The methods include Simple Differential Photometry (SDP),  \texttt{quaver} \citep{2023arXiv231008631S}, \texttt{eleanor}'s 1D Postcard background subtraction \citep{2019PASP..131i4502F}, and all three methods within the \textit{Corrector Class lightkurve} package (Regression, Cotrending Basis Vector, and Pixel Level Level De-correlation).
\begin{figure}[ht]
    \centering
    \includegraphics[scale=0.5, angle =0]{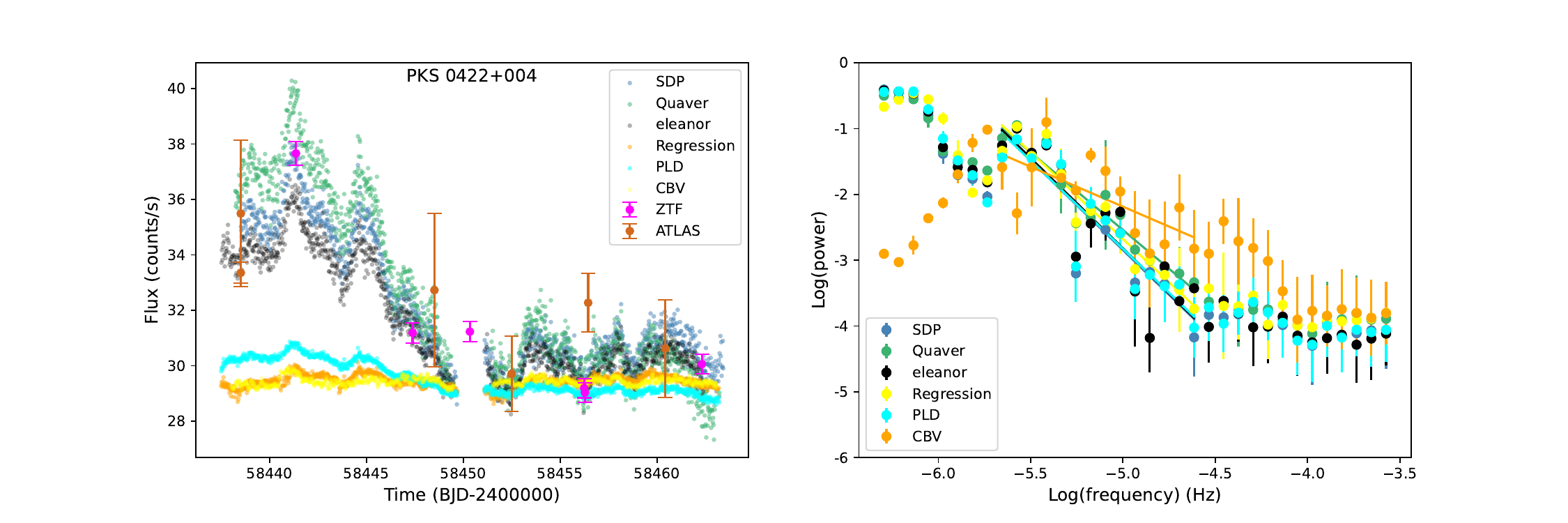}
    \caption{Comparison of all 6 light curve extraction methods and resulting PSDs for the blazar PKS~0422+004.}
    
    \label{fig:allmethods}

\end{figure}
\FloatBarrier
\subsection{Simple Differential Photometry}
    \label{sec:sdp}
    
The SDP method utilizes the principles of standard differential photometry with \textit{TESS.cut}, and routines from the \texttt{lightkurve} package.  It provides a background subtraction that utilizes pixels in each individual cutout of the target. Unlike the other methods we tested, it is interactive and does not function as an automated reduction pipeline. The steps are as follows:

\begin{itemize}

    \item{Using \textit{TESS.cut}, a $30\arcmin \times 30\arcmin$ cutout of the desired target is downloaded. Any data with non-zero quality flags are deleted.}

    \item{The \textit{tpf.interact()} command from the \textit{lighkurve} package is used to select ten pixels with the lowest count rates within the cutout (avoiding pixels near the edge of the cutout) to measure the background. The background value per pixel for each cutout is taken as the average of these ten pixels.  While none of our chosen source cutouts had a strong gradient in the background, it is worth noting that this method to determine the background will underestimate it in such cases.}

    \item{Next, the target blazar and two presumed non-variable comparison stars of similar brightness within the cutout are identified; if possible, these are known comparison stars from the customary blazar finding charts \citep[e.g.,][]{1996A&AS..116..403F,1985AJ.....90.1184S}.}

    \item{The brightest pixel at the location of the target, plus any adjacent pixels that are 50 cts s$^{-1}$ or higher than the background, are included in the photometric aperture. 
    Note: Apertures for the comparison stars and target were created separately using this process and they were not necessarily the same size. The fluxes of the target and each comparison star in the apertures are corrected for the background, using the previously determined background estimate.}
    \item{A ratio of the flux of the comparison stars is determined and plotted to confirm that the comparison stars are non-variable. Ratios of the fluxes of the blazar to each comparison star are then constructed and plotted as the blazar light curve, yielding a normalized light curve.}
    
\end{itemize}   

\subsection{\texttt{Quaver}}
    \label{sec:quaver}

\texttt{Quaver} \citep{2023arXiv231008631S} is a python-based TESS data reduction pipeline designed specifically for AGN.  Its development was funded by the TESS Guest Investigator program (PI: K.~L. Smith). Spacecraft systematics, scattered background light and variability from contaminating sources are modeled and removed using a principal component analysis.  

\texttt{Quaver} analyzes faint background pixels to model additive effects such as scattered light. It separately models multiplicative effects utilizing bright stars to determine effects from spacecraft systematics and source contamination. \texttt{Quaver} provides three different reduction methods: full hybrid, simple hybrid, and simple reduction using principal component analysis. Under the operation of this pipeline, all potential methods are run simultaneously, and three light curves are produced with the same aperture and masking. The simple hybrid method is found to better match ground based data, but may not be able to distinguish and remove all signals from background or systematics. The ``full" method is more capable of handling these effects but may remove real variability, and may not scale as well to ground-based data. This pipeline and the user guide have been made publicly available\footnote{\url{https://github.com/kristalynnesmith/quaver}}. 

After selecting the preferred method, the user enters the target name or coordinates. \texttt{Quaver} provides the user with the currently available TESS cycle/sectors for the blazar. The user selects the cycle or sector from which to extract the light curve and \texttt{quaver} presents an TESScut postage stamp of the target, and the user identifies individual pixels for the aperture. We used the same aperture chosen for the Simple Differential Photometry method for our comparison. \texttt{Quaver} then analyzes the data from the chosen aperture through its matrix, locating additive and multiplicative components, allowing the user to mask out cadences with large detected systematics. The outputs include a panel showing the corrected light curve, the components in the design matrices, and the extraction aperture. The output units are cts~s$^{-1}$. 

\subsection{\texttt{eleanor}}
    \label{sec:eleanor}
    
The \texttt{eleanor} \footnote{\url{https://github.com/afeinstein20/eleanor}} package \citep{2019PASP..131i4502F} is a TESS data reduction pipeline that allows the user to extract and systematically correct light curve data for any given target within a TESS FFI. It was also funded via the TESS Guest Investigator program (PI: A.~Feinstein). This pipeline uses a pre-determined aperture selection, with the ability to customize the code and use intermediate data products to select a custom aperture. For this study, the predetermined aperture was used for all targets. We determined that the use of a user defined aperture did not effect the final light curve or its behavior. Similar to  \texttt{quaver}, \texttt{eleanor} has multiple options to correct the background: 1D postcard background subtraction, 1D TPF background subtraction, and 2D TPF background subtraction. For this study, the 1D postcard background subtraction was used for all targets. The output units are normalized flux.

\subsection{Regression}
    \label{sec:regression}
This is one of the methods available as part of the \texttt{lightkurve} python package to remove systematic noise trends in the TESS data. Pixels from outside the user-defined photometric aperture are used to construct a design matrix of vectors that model systematic noise, scattered light, and thermal ramping within the FFI. These vectors are designed to model the systematic noise and are then used to remove this noise from the uncorrected light curve. For this study we followed the recommendation to apply five principal components while using principal component analysis when picking the number of vectors with which we detrended each light curve. The aperture selection within this method does not allow for individualized pixel selection, and instead implements a grid aperture such as a $2 \times 2$ or $2 \times 3$ pixel aperture. The output units are cts~s$^{-1}$.

\subsection{Pixel Level De-correlation}
    \label{sec:pxl_decorr}
Also available via \texttt{lightkurve} is the Pixel-Level De-correlation (PLD) method. The PLD method is similar to the regression method, both in operating procedure and reduction method. It models the systematics induced by scattered light and spacecraft motion in pixels adjacent to the target by finding common trends in these pixels. These trends are then combined and removed via linear regression. Unlike the regression method, every pixel that is not the within the target aperture and is not another bright source is used to identify systematic trends. However, in this pipeline there is no option to select an aperture and instead the pipeline uses a predetermined mask for the target, background, and a PLD aperture mask. The user does have the option of how many principal components to use for the principal component analysis. For consistency with the regression method, 5 principal components were used for each target. The output units are cts~s$^{-1}$. 

\subsection{Cotrending Basis Vectors}
    \label{sec:cbv}
The third method available via the \texttt{lightkurve} package makes use of Cotrending Basis Vectors (CBV), similar to techniques used for \textit{Kepler} and K2 data \citep{2014PASP..126..948V}. CBVs are created for each sector and CCD for a TESS observation as part of the PDC pipeline, and are built from the most common systematic trends apparent in each observation. In total, there are three generic CBV sets created for each observation for each CCD: Single-scale, Multi-scale, and Spike. Single-scale applies a correction for the most common systematic trends in a single set of basis vectors, while multi-scale applies different corrections corresponding to different bandpasses of the systematic trends. Spike only contains and corrects for short impulsive spike systematics such as single readout events.

For this study, Single-scale and Spike CBV corrections were used. After applying the corrections a final ``Goodness metric" fit can be calculated to determine if the corrected light curve was over-fitted or under-fitted. The output units are cts~s$^{-1}$.
This ``goodness metric" check evaluates the modelling and subtraction of background systematics from the light curve via a over-fitting and under-fitting metric calculation. The over-fitting metric is calculated by an examination of the periodigram before and after the correction, and is looking for the injection of broad-band noise into the corrected light curve. The under-fitting metric is calculated by examining the correlation between the corrected light curve and other neighboring Science Processing Operations Center (SPOC) SAP light curves from the same observation. An ideal fit of the background and spacecraft systematics can be achieved by balancing these two metrics. For our purposes, the ``goodness metric" was used for the CBV method in order to achieve the best possible reduction of the corrected light curve. Nonetheless, this still led to light curves that did not display variability features similar to those seen in the remaining five methods.

\section{Ground-based Observations}
    \label{sec:gb_obs}

TESS was specifically designed to be a ``forward facing" mission, meaning contemporaneous ground based observations could be obtained of the same fields being observed by TESS. We utilized the publicly available databases from the Zwicky Transient Facility (ZTF) and the Asteroid Terrestrial-impact Last Alert System (ATLAS) as sources for contemporaneous ground truth observations.

\subsection{Zwicky Transient Facility}
    \label{sec:ztf}

ZTF \citep{2019PASP..131a8002B} observes the northern hemisphere sky with a cadence of approximately 2 days, utilizing the 48-inch Samuel Oschin Telescope (Schmidt-type) at the Palomar Observatory. The science camera is a 47-square-degree, 600 megapixel cryogenic CCD mosaic and observations are obtained in ZTF-specific \textit{g, r} and \textit{ i} filters. The data archive is publicly available via the NASA/IPAC Infrared Science Archive\footnote{\url{https://irsa.ipac.caltech.edu/Missions/ztf.html}}. We used the project-provided magnitudes for all but one of our blazars (3C~371). In the case of 3C~371, we requested the processed images and performed differential aperture photometry due to the presence of a strong underlying galaxy in this source (see Section \ref{sec:3c371} for further details). To best complement TESS's bandpass sensitivity of 600nm--1000nm, the red filter (ZTF$\_$r) data  was used for our comparison with the TESS observations. The ZTF$\_$r filter is a custom r filter supplied by Materion Precision Optics and spans 700nm--900nm \citep{2019PASP..131f8003B}. The number of data points varies from target to target, as will be seen in Section \ref{sec:results}, where one can compare 1ES~1011+496 and its plethora of simultaneous ZTF data to the few points acquired for targets such as PKS~0422+004. ZTF$\_$r magnitudes, $m$, were converted to flux, $f$, via the usual relationship between magnitude and flux, $m=-2.5\log (f)$. Using the \texttt{quaver}-created light curve, the offset between each ZTF data point and two  \texttt{quaver} flux measurements (one directly before and another directly after, typically separated by about a day) was found. The average of the offsets for all ZTF-\texttt{quaver} matched data points was then used as the multiplicative scale factor for the ZTF data  points to compare them with the \texttt{quaver} light curve. The remaining five methods' light curves we tested were then scaled by an empirically determined multiplicative scale factor to the \texttt{quaver} light curve for ease in comparison of the overall light curve shapes.

\subsection{Asteroid Terrestrial-impact Last Alert System }
    \label{sec:atlas}

ATLAS \citep{2018PASP..130f4505T} is comprised of 4 telescopes (2 located in Hawaii, 1 in Chile, and 1 in South Africa) surveying the sky for moving objects. It can also detect other, stationary transients. A forced photometry service is available, as is access to the calibrated CCD images\footnote{\url{https://fallingstar-data.com/forcedphot/}}.  Forced photometry is preformed by calculating a point-spread-function for each image based on high signal to noise stars, and a profile fit is forced at the target coordinates. However, we found in some cases the forced photometry returned negative magnitudes and fluxes for objects. We chose to download the raw frames and perform differential aperture photometry on the blazars in this study. The published comparison sequences from \citet{1996A&AS..116..403F}, \citet{1985AJ.....90.1184S}, \citet{1983AJ.....88.1301M}, as well as comparison sequences from the SMARTS Optical/IR Observations of blazars found online\footnote{\url{http://www.astro.yale.edu/smarts/glast/home.php$\#$}} were used. For sources without published comparison stars (1246+586, ON~246, and TXS~0506+056) we determined a differential magnitude for the source from comparison stars of similar magnitude in the CCD image. The resulting magnitudes were converted to fluxes via the usual relationship between magnitude and flux. 

In the same manner as we used for ZTF, we empirically determined a multiplicative scale factor to convert the fluxes derived from the ATLAS magnitudes to the \texttt{quaver} units of cts s$^{-1}$.  As can be seen in Figures \ref{fig:allmethods} and \ref{fig:alltargets}, the ATLAS error bars (3$\%$ or more) are typically larger  than those for ZTF, especially for the fainter sources, due to the smaller sizes of the ATLAS telescopes (0.5m) compared to the 48-inch Oschin telescope.  Hence, ATLAS is most useful for the brighter blazars.

\section{Results}
    \label{sec:results}

The TESS observations used for all sources were 30-minute cadence FFI observations. A common characteristic of TESS light curves is the $\sim$1 day gap in the middle of each sector due to the pause in observations and spacecraft repointing for data download. In addition, thermal ramps lasting about one day are often recognizable by their ``hook'' shape at the beginning of an observation. The different methods of light curve extraction that we are comparing each have their own recipes for removing such effects. Consequently, we find the usable length of the light curves as well as the lengths of the data download gaps may  differ between sources and also depend on the extraction method.  The TESS sectors differ somewhat in length as well. The ``Science Days" column in Table 2 lists the duration of the actual data recording for each sector, as given in the TESS Cycle 1 and Cycle 2 Data Release Notes\footnote{\url{https://heasarc.gsfc.nasa.gov/docs/tess/cycle1_drn.html}}. 

As an example, the left panel of Figure \ref{fig:allmethods} displays the extracted light curves for all six methods, along with the ground based (ZTF and ATLAS) data for one of our targets, PKS~0422+004. As described earlier, the light curves were scaled to the  \texttt{quaver} counts s$^{-1}$ for comparison. The right panel shows the resulting PSDs which are constructed as described in \citet{2023ApJ...951...58W}. Briefly, the PSD is found via a periodogram analysis of the time series, and then binned in steps of 0.08 in log($\nu$). The PSD is normalized, hence its units are dimensionless. A linear fit to the PSD is performed as described below. From simulations of 1000 light curves with the TESS cadence and duration and gap, generated via the method described in \citet{1995A&A...300..707T}, with an input PSD slope of $-2.0$ we found that the data gap did not significantly effect our PSD slope determination. See Appendix A for further details.

A key result is that the PLD, CBV, and regression methods all produced light curves with much lower variability amplitudes than the ground based data and the other light curve extraction methods. In three cases (ON~325, OQ~530 and PKS~1424+240), the CBV method produced anomalous results. The automated CBV method failed to subtract the background flux from the corrected light curve for unknown reasons and the resulting CBV light curve had no resemblance to those produced by the other 5 methods. In addition, the flux decline which dominates the first half of the TESS observation of PKS~0422+004 is eliminated, as seen in Figure \ref{fig:allmethods}.
In general, we found that while the light curves produced from the PLD, CBV, and regression methods do retain the short term character of the variability, they both squash the amplitude and significantly diminish any long term trends.  Changing the parameters employed in these methods (number of regressors, number of CBVs,  etc.) did not improve the corrected light curves. For the reminder of our analysis, we will focus on the the three methods (\texttt{quaver}, \texttt{eleanor}, and SDP) which produced light curves consistent with the ground truth observations.

Comparisons between the SDP,  \texttt{quaver}, and \texttt{eleanor} methods, and the ground based photometry for each source, are displayed in the left panels of Figure \ref{fig:alltargets}. The right panels display the PSDs for the light curves from each of the three extraction methods. Table \ref{tab:TESS_PSD} summarizes the slope of the PSDs determined from each extraction  method's light curves, along with the range of frequencies within which the PSD was calculated. The frequency range for the power law fit was chosen to be from the point where the power law clearly emerges from the white noise portion of the PSD seen at higher frequencies (log($\nu$) = -4.6), to the frequency where the PSD no longer displayed a power law nature (typically log($\nu$) = -5.9, except where noted in Table 3 and discussed below). The PSDs are binned to 0.08 in log frequency prior to fitting and the error on the slope is the error from the linear least squares fit. Overall, we find that the ground based observations confirm the character of the TESS observations as described and evaluated below, and  the three different extraction methods compared here produce similar overall morphology of the variability. 

Multiple attributes of the light curves, i.e., the Coefficient of Dispersion, a least squares fit test between each target and the ZTF data, and the $\frac{Max}{Min}$ ratio for fractional variability, were evaluated for each of the three methods. Our findings are presented in Table \ref{tab:TESS_stats}.
An evaluation of the coefficient of dispersion was conducted by taking the standard deviation and mean of the cts~s$^{-1}$ during quiescent periods of each light curve. These periods would range from one to three days of TESS observations. Once the standard deviation ($\sigma$),  and mean ($\bar x$), for these quiescent periods were found, the coefficient of dispersion was calculated as the ratio $\sigma/{\bar x}$. As seen in Table \ref{tab:TESS_stats}, \texttt{eleanor}-produced light curves had the lowest coefficient of dispersion for the majority of targets and \texttt{quaver} -produced targets had the most instances of the highest coefficient of dispersion.

We evaluated the quality of fit between the ground based data and the corrected light curves by performing a least squares fit. This was done by finding the best optimized scale factor for the simultaneous ZTF data for each method's light curve. An initial scale factor was created by finding the ratio of each nearest light curve point to each ZTF data point. This scale factor was then optimized to best fit each data point by minimizing a $\chi^2$ function. These $\chi^2$ values for each target using each method are presented in Table \ref{tab:TESS_stats}. Statistically, \texttt{eleanor} best matched the simultaneous ZTF data in most cases. Explicitly, the $\chi^2$ value was the smallest for \texttt{eleanor} for seven targets, followed by \texttt{quaver} for three, and SDP for one. However, the optimized scalar was designed to only take the nearest points to the ZTF data in consideration and not the average dispersion of light curve points around the ZTF data. Therefore, the light curves and ground-based data remained scaled to the \texttt{quaver} produced light curve for the best visual comparison.

The $\frac{Max}{Min}$ ratio of the cts~s$^{-1}$ was calculated to observe the overall variability within each light curve as evaluated by the different methods.  For the blazar PKS 0422+004, we calculated the $\frac{Max}{Min}$ ratio for all six methods; however, for the rest of the targets we only found those ratios for the light curves produced by the SDP, \texttt{quaver}, and \texttt{eleanor} methods. As seen in Table \ref{tab:TESS_stats}, PKS 0422+004 showed a substantial range of variability (between 1.30 and 1.47) when using SDP, \texttt{quaver}, or \texttt{eleanor}. On the other hand, the calculated $\frac{Max}{Min}$ ratios for the \texttt{Lightkurve Corrector Class} packages were the following: Regression = 1.07; PLD = 1.08; and CBV = 1.03. We thus quantify the extent of the visually obvious squashing of the blazar light curves produced by the \texttt{Lightkurve Corrector Class} reduction methods. 

The final two columns of Table \ref{tab:TESS_stats} are the average and standard deviation of the flux in cts~s$^{-1}$ over the entire observation of the target. These values are derived from the photometric aperture used in the SDP reduction, and are simply the counts within the aperture after the background estimate is subtracted
..
\begin{figure}[ht]

\begin{center}
\includegraphics[scale=.85,page=1,angle =0]{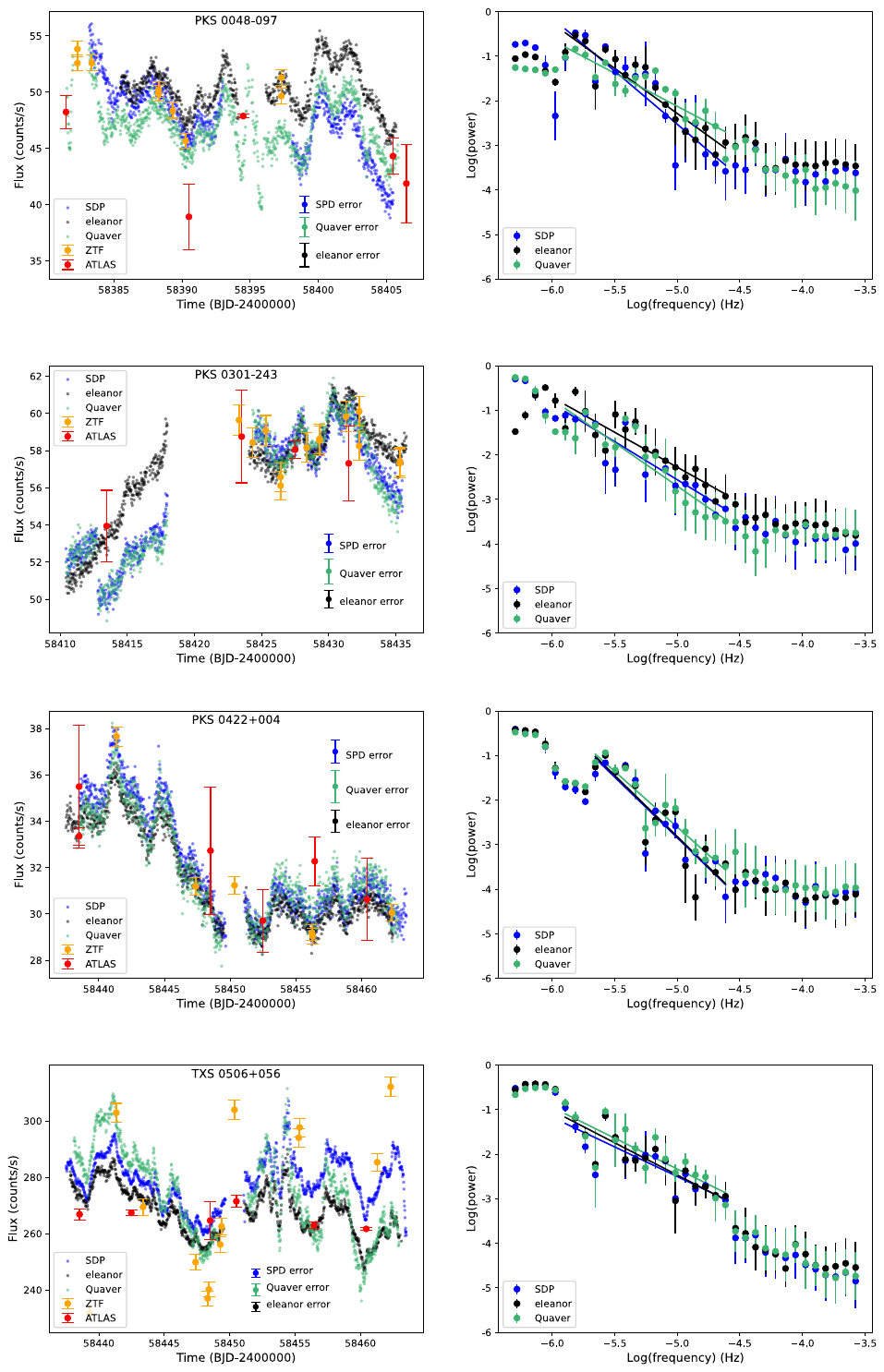}
\caption{The left panels show the comparison of the SDP, \texttt{eleanor} and  \texttt{quaver} light curve extraction methods, along with the ground based observations. The right panels show the PSDs and linear fits to them for those three extraction methods.}
\label{fig:alltargets}
\end{center}
\end{figure}
\newpage
\begin{figure}

\begin{center}
\includegraphics[scale=.85,page=2,angle =0]{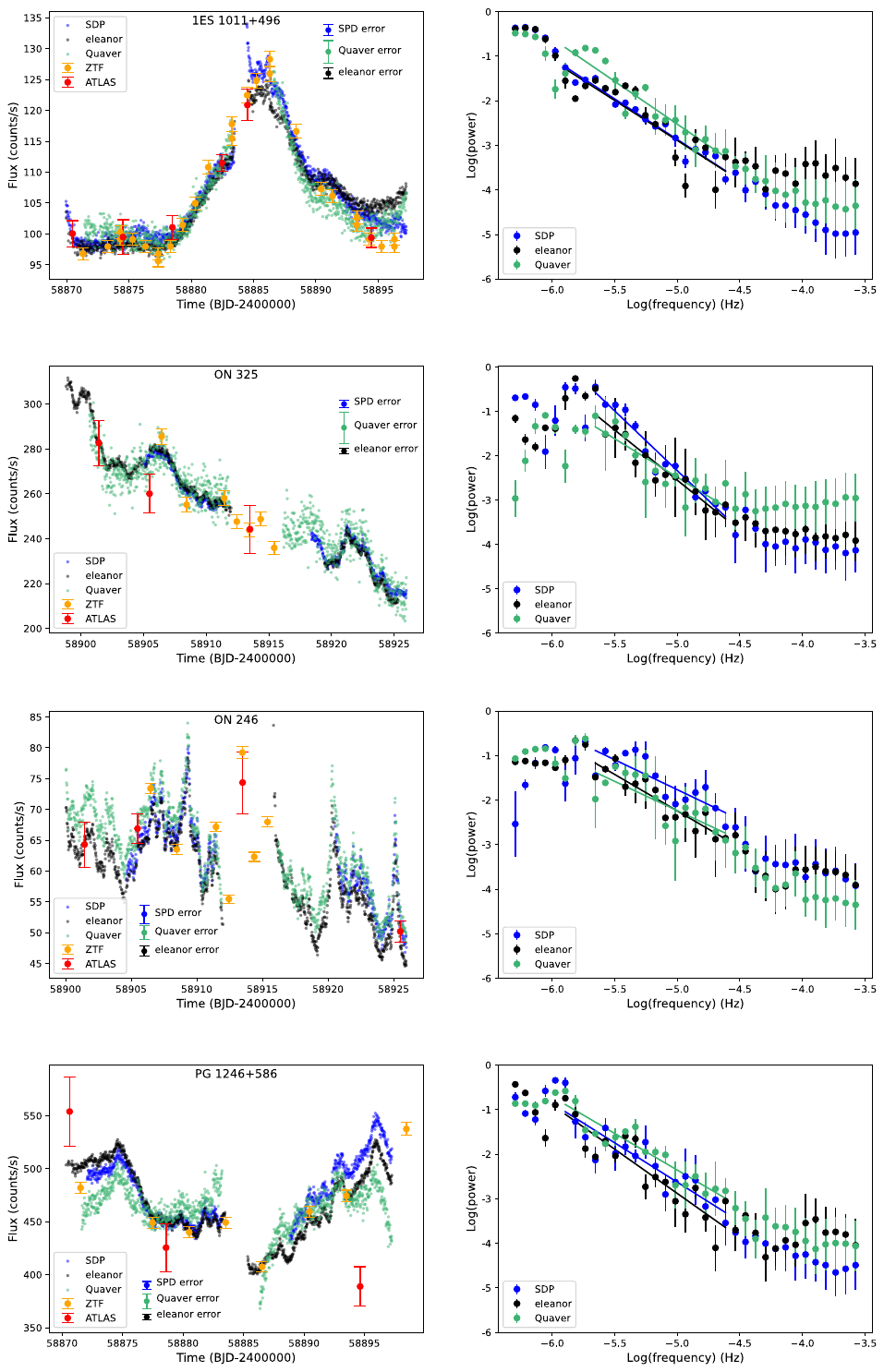}
\addtocounter{figure}{-1} 
\caption{continued.}
\newpage
\end{center}
\end{figure}
\begin{figure}

\begin{center}
\includegraphics[scale=.85,page=3,angle =0]{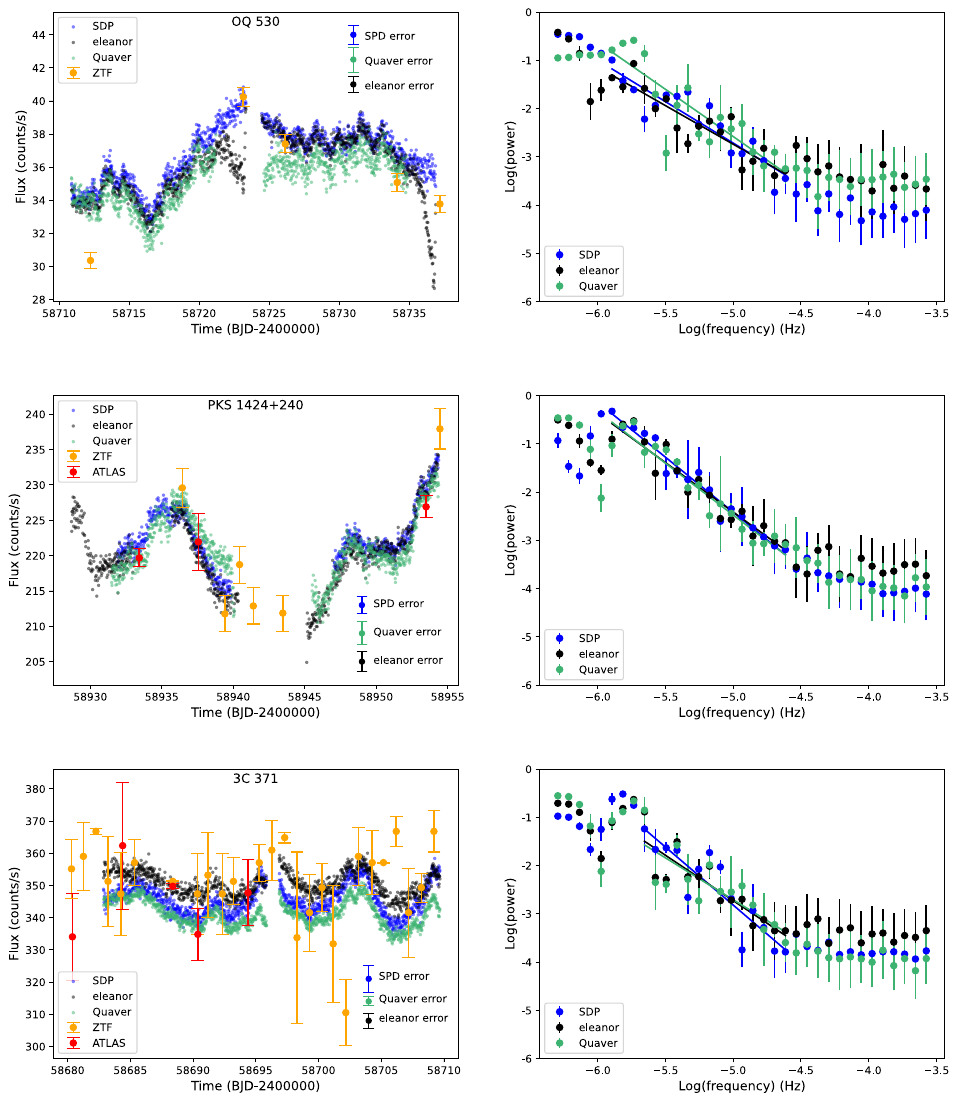}
\addtocounter{figure}{-1} 
\caption{continued.}
\end{center}
\end{figure}

In order to quantify our conclusion from the visual inspection of light curves that the three different extraction methods are in good agreement, we compared the PSDs and their slopes for each extraction method. We chose the PSD for this comparison because it is one of the most widely used methods to characterize blazar light curves, though it has known limitations \citep{2001MNRAS.323L..26U}. We calculated an average slope error for each object from the three different methods and used that as the basis to compare the methods for each source. We find that for 8 of the 11 sources, the slopes agree to within two standard deviations of the average error. The outliers are PKS~0048$-$097, ON 325 and 3C 371. We note that for these targets, \texttt{quaver} yields a shallower slope compared to the PSD's from SPD and \texttt{eleanor}.  Inspecting the light curves of these three sources, \texttt{quaver} includes sampling the earliest into the observation (for PKS~0048$-$097) and more points near the data download gap when compared to SDP and \texttt{eleanor} than any of the other sources. This additional data in the light curve may be introducing power at lower frequencies, around $10^{-5}$Hz,  thus resulting in the shallower slope of the PSD.

\subsection{Notes on individual sources}
    \label{sec:notes}

In this section, we comment on three sources (3C~371, TXS~0506+056 and OQ~530) that had unique characteristics or challenges.  

\subsubsection{3C~371}
    \label{sec:3c371}
The ground based observations of 3C~371 in general follow the shape of the TESS observations (Figure \ref{fig:alltargets}). 3C~371 has a conspicuous galaxy component in ground based observations which can affect the ground based photometry, especially in cases of poor seeing. There are several observations which differ significantly from the overall light curve; these are likely points effected by the combination of poor seeing and the underlying galaxy component. The ZTF observations presented were manually re-reduced utilizing differential aperture photometry and the comparison stars recommended by the Whole Earth Blazar Telescope (\url{https://www.oato.inaf.it/blazars/webt/3c-371-ugc-11130-s4-180769-4c-69-24-txs-1807698/}) in order to minimize such effects.

\subsubsection{TXS~0506+056} 
    \label{sec:TXS0506+056}
Similar to 3C~371, the ground based observations mostly agree with the long term behavior of the TESS observations, with some discrepant points (Figure \ref{fig:alltargets}). Unlike 3C~371, TXS~0506+056 does not have a significant galaxy component. We have manually re-reduced the ZTF observations utilizing differential aperture photometry with comparison stars of similar magnitude on the ccd frame (no published comparison sequence for TXS~0506+056 exists). The discrepancy between the TESS and ground based points (which are primarily in the second half of the TESS observation), is currently unresolved. However, this does not affect our conclusion that the TESS and ground based observations are in general agreement. 

\subsubsection{OQ~530}
    \label{sec:OQ530}
As seen in Figure \ref{fig:alltargets}, the overall morphology of the \texttt{quaver} and SDP light curves are confirmed by the ground based data. Toward the end of the first observing segment, the \texttt{eleanor} light curve deviates significantly from both  \texttt{quaver} and SDP-produced light curves.  At the beginning of the second segment, the  \texttt{quaver} light curve is much flatter than the SDP or \texttt{eleanor} light curves; this could be the result of  \texttt{quaver} over correcting the observed flux decline, by interpreting it as a background variation rather than astrophysical variability. A similar effect likely occurred with the \texttt{eleanor} light curve in the time period just before the gap in the light curve due to the data download.

\subsection{Inter-day and intra-day (microvariability) comparison}
    \label{sec:inter_intra_comp}

\begin{figure}

\vspace{0mm}
\begin{center}
\includegraphics[trim=0cm 5cm 0cm 0cm,clip, scale=0.5, angle =0, scale = 1.5]{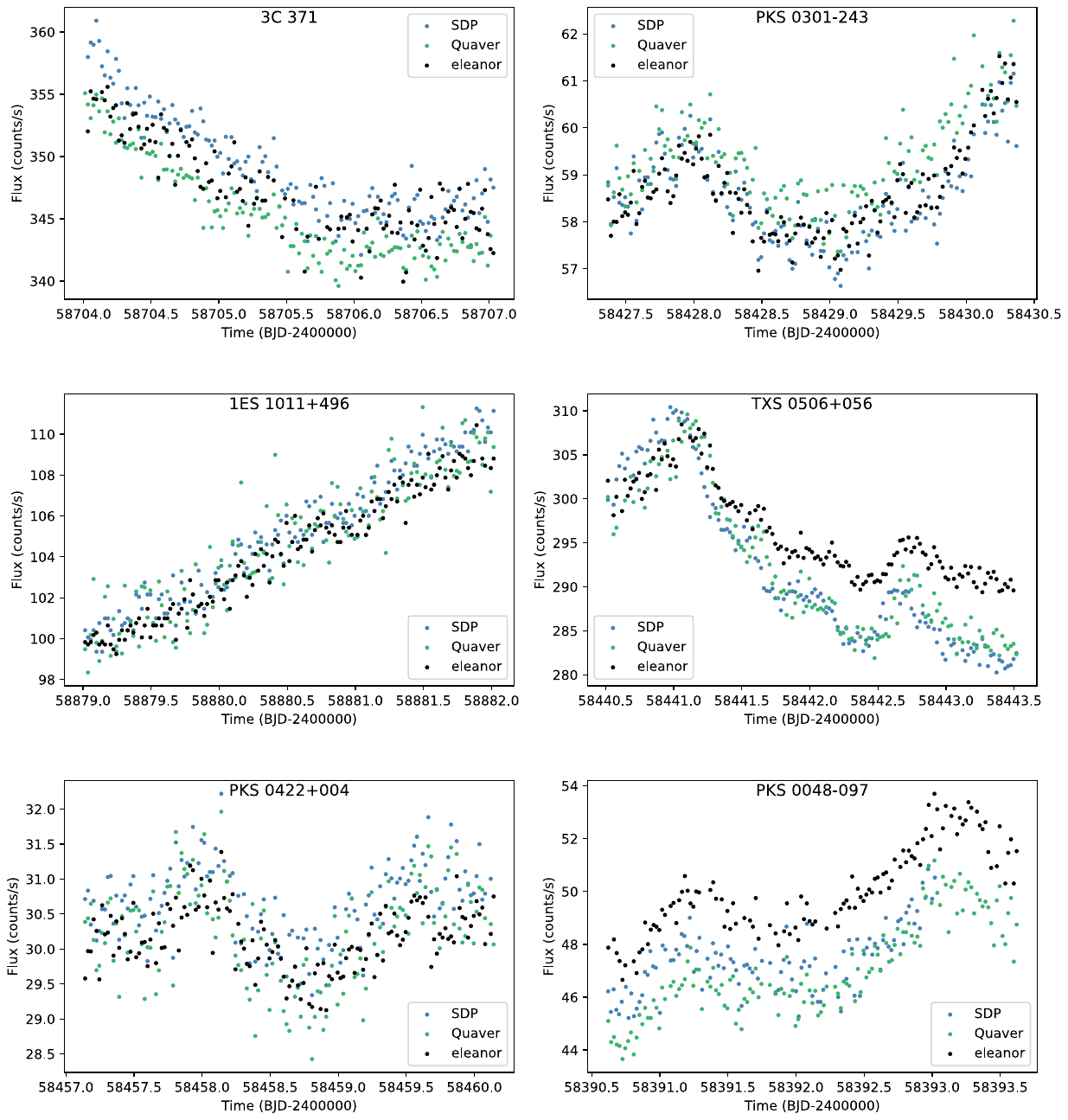}
\caption{Examples of the comparison of inter- and intra- day variability from our sample for the SDP, \texttt{eleanor} and  \texttt{quaver} extraction methods}
\vspace*{-300mm}

\label{fig:shorttermvar}
\end{center}
\end{figure}

In addition to comparing how well the longer term (day to weeks) variability is recovered, we also examined how well the more rapid (hours to days) variations in our 30-min FFI data are recovered via our three principal extraction methods (SDP,  \texttt{quaver} and \texttt{eleanor}). We did not have any ground-based observations to validate these TESS observed inter- and intraday (microvariability) variations. Figure \ref{fig:shorttermvar} displays examples of 3 day segments of the extracted light curves from some of our blazars, each highlighting rapid variability with different morphology. We see that in all cases, the extraction methods agree down to the most rapid variations resolvable by the TESS 30-min FFI observations. 
\newpage
\FloatBarrier
\FloatBarrier
\section{Conclusions}
    \label{sec:conclusions}

Our main conclusions are:
\begin{enumerate}
\item{\texttt{Quaver}, \texttt{eleanor}, and SDP all reproduce TESS lightcurves with the overall variability amplitudes and trends indicated by the ground truth observations. }

\item{The noise removal methods available as part of the corrector class in the \texttt{lightkurve} package, which are methods adapted from the \textit{Kepler}/K2 mission (Regression, PLD, and CBV) did not accurately reproduce the variability in the blazar targets. Long term trends were mistaken for systematics and removed and short timescale variability amplitudes were significantly reduced.}

\item{\texttt{Quaver},  \texttt{eleanor}, and SDP all produce similar light curves, down to the most rapid timescales measured. Unsurprisingly, the resulting PSDs indicate the same variability characteristics and can be fit with power laws of very similar slopes.   Explicitly, the three method's PSD slopes for 8 of the 11 blazars   agreed to within $2 \sigma$. }

\item{There is no single  ``right" method to extract blazar light curves from TESS.  \texttt{Quaver} is the most configurable of the three successful methods; however, it notably also has the largest dispersion, as observed in the light curves of 1ES~1011+496, ON~325, and PG~1246+586. \texttt{Eleanor} is more of a black box and other studies have pointed out it is very sensitive to the aperture selection \citep{2021MNRAS.501.1100R}. SDP in its current implementation is a process, rather than a package, and is the most time intensive of the three successful methods. 
\item{Overall, for a user who has a large survey of targets that were observed in multiple TESS Cycles, \texttt{quaver} allows for the quickest and most configurable reduction. For single observations, either \texttt{eleanor} or \texttt{quaver} should work well.}}
\end{enumerate}

Running any of these extraction methods in a pipeline fashion with little or no validation may well  produce light curves that may be either over- or under-corrected and could lead to misleading results. Ultimately, the user must critically evaluate the results of their extraction method, and, whenever possible, validate the results with ground truth observations.  The three methods that work successfully on highly variable blazar light curves are also likely to work well on other variable extragalactic sources such as non-blazar AGN, supernova, tidal disruption events, etc., whose light curves that can be extracted from TESS data. We note that Michael Fausnaugh kindly ran a subset of our sources through the \texttt{ISIS} software \citep{2021ApJ...908...51F}, and the resulting light curves were qualitatively very similar to the light curves produced by quaver, SDF and eleanor.

All the {\it TESS} data used in this paper can be found in MAST: \dataset[10.17909/x9x0-br63]{http://dx.doi.org/10.17909/x9x0-br63}
\section{Acknowledgements}
\label{sec:acknowledgements}
The material is based upon work supported by the NASA National Space Grant College and Fellowship Program and the Kentucky Space Grant Consortium under NASA award number 80NSSC20M0047. We thank the anonymous reviewer for useful comments that led to improvements in the manuscript. We would also like to thank Dr. Krista Lynne Smith for early access and insight to her \texttt{quaver} data reduction method \url{https://github.com/kristalynnesmith/quaver}. This work made use of publicly available Zwicky Transit Facility (ZTF) and Asteroid Terrestrial-impact Last Alert System (ATLAS) data. ZTF is funded in equal part by the US National Science Foundation and an international consortium of universities and institutions. Its data can be found within the NASA/IPAC Infrared Science Archive \url{https://irsa.ipac.caltech.edu/Missions/ztf.html}. ATLAS is funded by NASA, and developed and operated by the University of Hawaii's Institute for Astronomy and its data can be found at \url{https://fallingstar-data.com/forcedphot/}.
Multiple software packages were used during our data analysis of the TESS light curves. These packages include: Astrocut \citep{2019ascl.soft05007B}, a package that creates the same cutout across all FFIs that share a common pointing to create a time series of images on a small portion of the sky; Astropy \citep{2013A&A...558A..33A},  a common core package for Astronomy in Python; Astroquery \citep{2019AJ....157...98G},  a set of tools for querying astronomical web forms and databases; Lightkurve \citep{2018ascl.soft12013L}, a Python package for Kepler and TESS data analysis.
\facility{ATLAS, TESS, \& ZTF}
\software{Astrocut \citep{2019ascl.soft05007B}, Astropy \citep{astropy:2013, astropy:2018, astropy:2022}, Astroquery \citep{2019AJ....157...98G}, eleanor \citep{2019PASP..131i4502F}, quaver \citep{2023arXiv231008631S}, \& Lightkurve \citep{2018ascl.soft12013L}} 
%
%

\bibliographystyle{aa}
\bibliography{bibliography}

\begin{deluxetable}{ccccccccc}
        \label{tab:TESS_PSD}
\tablecaption{Power Spectral Densities (PSD)}
\tablewidth{0pt}
\tablecolumns{10}
\tabletypesize{\small}
\tablehead{
         \colhead{Name}
         &\colhead{Fit Range}
         &\colhead{Fit Range}
         & \colhead{SDP}
         & \colhead{ \texttt{quaver}}
         & \colhead{\texttt{eleanor}}   
         & \colhead{Regression}
         & \colhead{PLD}
         & \colhead{CBV}
        \\
         \colhead{}
         & \colhead{Log(Hz)}
         & \colhead{(days)}
         & \colhead{}
         & \colhead{}
         & \colhead{}
         & \colhead{}
         & \colhead{}
         & \colhead{}
        \\
         }

\startdata
PKS~0048-097 & -5.9 -- -4.6 & 9.19 -- 0.46 &	-2.39±0.27 & -1.48±0.18 & 	-2.02±0.21 &	-2.66±0.14 & 	-2.42±0.15 & 	-2.08±0.19\\
PKS~0301-243 & -5.9 -- -4.6 & 9.19 -- 0.46 & 	-1.71±0.22 & 	-1.94±0.23 & 	-1.57±0.19 & 	-1.58±0.36 & 	-2.16±0.24 & 	-1.25±0.16\\
PKS~0422+004 & -5.7 -- -4.6 & 10.80 -- 0.46 & 	-2.72±0.34 & 	-2.36±0.20 & 	-2.74±0.40 & 	-2.63±0.21 & 	-2.68±0.33 & 	-1.21±0.38\\
TXS~0506+056 & -5.9 -- -4.6 & 9.19 -- 0.46 & 	-1.33±0.21 & 	-1.38±0.22 & 	-1.45±0.20 & 	-1.22±0.21 & 	-1.51±0.21 & 	-0.94±0.19\\
1ES~1011+496 & -5.9 -- -4.6 & 9.19 -- 0.46 & 	-1.82±0.10 & 	-1.92±0.18 &  -1.79±0.26 & 	-1.99±0.24 & 	-2.21±0.18 & 	-1.43±0.25\\
ON~325 & -5.7 -- -4.6 & 10.80 -- 0.46 & -2.69±0.18 &  -1.73±0.28 & -2.26±0.23 & -2.00±0.29 & 	-2.11±0.24 & \nodata\\
ON~246 & -5.7 -- -4.6 & 10.80 -- 0.46 &	-1.34±0.27 & 	-1.31±0.32 & -1.64±0.20 &	-1.23±0.24 & 	-1.29±0.35 & 	-0.41±0.07\\
PG~1246+586	& -5.9 -- -4.6 & 9.19 -- 0.46 & 	-1.77±0.21 & 	-1.64±0.16 & 	-1.99±0.24 & 	-1.60±0.29 & 	-2.38±0.13 & 	-2.01±0.20\\
OQ~530 & -5.9 -- -4.6 & 9.19 -- 0.46 & 	-1.72±0.20 & 	-1.96±0.29 & -1.60±0.19 & 	-1.26±0.22 & 	-1.93±0.14 & \nodata\\
PKS~1424+240 & -5.9 -- -4.6 & 9.19 -- 0.46 & 	-2.28±0.11 & 	-2.15±0.15 & 	-2.05±0.15 & 	-1.77±0.28 & 	-1.77±0.24 & 	\nodata\\
3C~371 & -5.7 -- -4.6 & 10.80 -- 0.46 &  -2.40±0.34 & 	-1.73±0.35 & 	-1.88±0.29 &  -1.96±0.27 & 	-2.42±0.26 & 	-1.49±0.21\\
\enddata
\end{deluxetable}

\begin{deluxetable}{ccccccccccccr}
        \label{tab:TESS_stats}
\tablecaption{Light Curve Statistics}
\tablewidth{0pt}
\singlespace
\tablecolumns{4}
\tabletypesize{\small}
\tablehead{
         \colhead{Name}
         & \colhead{SDP}
         & \colhead{\texttt{quaver}}
         & \colhead{\texttt{eleanor}}
         & \colhead{SDP}
         & \colhead{\texttt{quaver}}
         & \colhead{\texttt{eleanor}} 
         & \colhead{SDP}
         & \colhead{\texttt{quaver}}
         & \colhead{\texttt{eleanor}}
         & \colhead{Average}
         & \colhead{SD}
                 \\
         \colhead{}
         & \colhead{CoD}
         & \colhead{CoD}
         & \colhead{CoD}
         & \colhead{$\chi^2$}
         & \colhead{$\chi^2$}
         & \colhead{$\chi^2$}
         & \colhead{$\frac{Max}{Min}$}
         & \colhead{$\frac{Max}{Min}$}
         & \colhead{$\frac{Max}{Min}$}
         & \colhead{Counts}
         & \colhead{cts s$^{-1}$}
        \\
         \colhead{}
         & \colhead{$\frac{\sigma}{Mean}$}
         & \colhead{$\frac{\sigma}{Mean}$}
         & \colhead{$\frac{\sigma}{Mean}$}
         & \colhead{}
         & \colhead{}
         & \colhead{}
         & \colhead{}
         & \colhead{}
         & \colhead{}
         & \colhead{cts s$^{-1}$}
         & \colhead{}
        \\
         }

\startdata
PKS~0048-097 & 0.0214 & 0.0189 & 0.0143 &43.1& 39.1& 4.6& 1.44 & 1.29 & 1.22 & 52.3 & 3.1\\
PKS~0301-243 & 0.00965& 0.00954& 0.0126 & 24.1 & 18.2 &13.9 & 1.24 & 1.48 & 1.22 & 58.3 & 3.4\\
PKS~0422+004 &0.0143 & 0.0190 & 0.0144& 23.6 & 24.0 & 4.6 & 1.30 & 1.33 & 1.47 & 35.3 & 3.3 \\
TXS~0506+056 & 0.00465& 0.00713 & 0.00513 & 7410 & 7370 & 7260& 1.18& 1.21 & 1.17 & 745.4 & 23.1\\
1ES~1011+496 & 0.00858& 0.0172& 0.00802& 262 & 182 & 382 &1.38 & 1.32 & 1.30 & 110.0 & 8.7\\
ON~325 & 0.00630& 0.0267 & 0.00487 &108 & 81.2& 122 &1.32 &1.48 &1.48 & 249.9 & 20.3 \\
ON~246 & 0.0285&0.0193 &0.0167 &47 &32.5 & 21.1&1.71 &1.63 &1.88 & 68.0 & 7.3\\
PG~1246+586	& 0.00796 & 0.0145 & 0.00640 &484 & 1240& 325 & 1.28 & 1.38& 1.32 & 153.7 & 8.4\\
OQ~530 &0.0164 & 0.0220&0.0143 &51.9 &45.5 &121 &1.27 &1.24 & 1.38 & 36.9 & 1.7\\
PKS~1424+240 &  0.00552 &  0.00745 & 0.00673  &32.6 & 76.3 & 31.4 &1.09 & 1.10&1.14 & 231.0 & 3.9\\
3C~371 & 0.00615 & 0.00415 & 0.00658 & 4150 & 3750 & 3660 &1.06 &1.02 &1.06 & 331.2 & 15.4\\
\enddata
\end{deluxetable}
\pagebreak

\appendix{}
\section{Simulations of the effect of the TESS data download gap on the PSD}
    \label{sec:simulations}

TESS observations in a given sector contain an $\sim$ 1 day gap due to a pause in observations for data download. For the sectors used in this study, the gap ranged from 0.95 -- 1.35 days. However, the actual gap in the time series can be longer depending on the methodology used to extract the light curve. We examined the effect a gap in the middle of a simulated TESS observation of a blazar with an underlying PSD slope of $-1.5, -2.0$ and $-2.5$. We simulated 1000 light curves with cadence of 30 minutes and a duration of 28 days using the method of \citet{1995A&A...300..707T}. The PSD slope was measured for the light curve with no gap as well as for the light curve with a 1.2 day or 5 day gap, as described in the text. The gap of 1.2 days represents the average expected length of the gap for the sectors examined and 5 days the largest gap we saw in our 11 sources. 

Figure 4 shows the results for an input PSD slope of $-2.0$. The average output PSD slope was $-2.04 \pm 0.11$ for no gap, $-1.98 \pm 0.11$ for a gap of 1.2 days and $-1.97 \pm 0.12$ for a gap of 5 days. Quoted for each slope is the  error  on the linear fit to the binned PSD. We define the level at which a flattening or steepening of the PSD slope would be significant to be twice the average of the propagated errors from the difference between the PSD slope from the light curve with no gap and the light curve with a gap.

We found that a flattening of the PSD slope can occur in the time series with a gap of either 1.2 or 5 days. For a typical blazar PSD slope of $-2.0$, it occurs at a 2$\sigma$ significance  for 4.9\% of the light curves with a 1.2 day gap and 9.9\% of the light curves with a 5 day gap.  For an input PSD slope of $-1.5$, a flattening at greater than a 2$\sigma$ significance occurred in 1.2\% of the light curves with a 1.2 day gap and in 8.9\% of the light curves with a 5 day gap.  With an input PSD slope of $-2.5$, a flattening at a 2$\sigma$ or greater significance occurred in 14.9\% of the light curves with a 1.2 day gap and in 19.7\% of the light curves with a 5 day gap.

We conclude that the effect of a download gap in the TESS observation of up to 5 days has an negligible effect on our calculated PSD slopes. We also note it appears that flattening at a significant level increases both with the gap size and with steeper PSD slopes. 
\begin{figure}[ht]

\begin{center}
\includegraphics[scale=.85,page=1,angle =0]{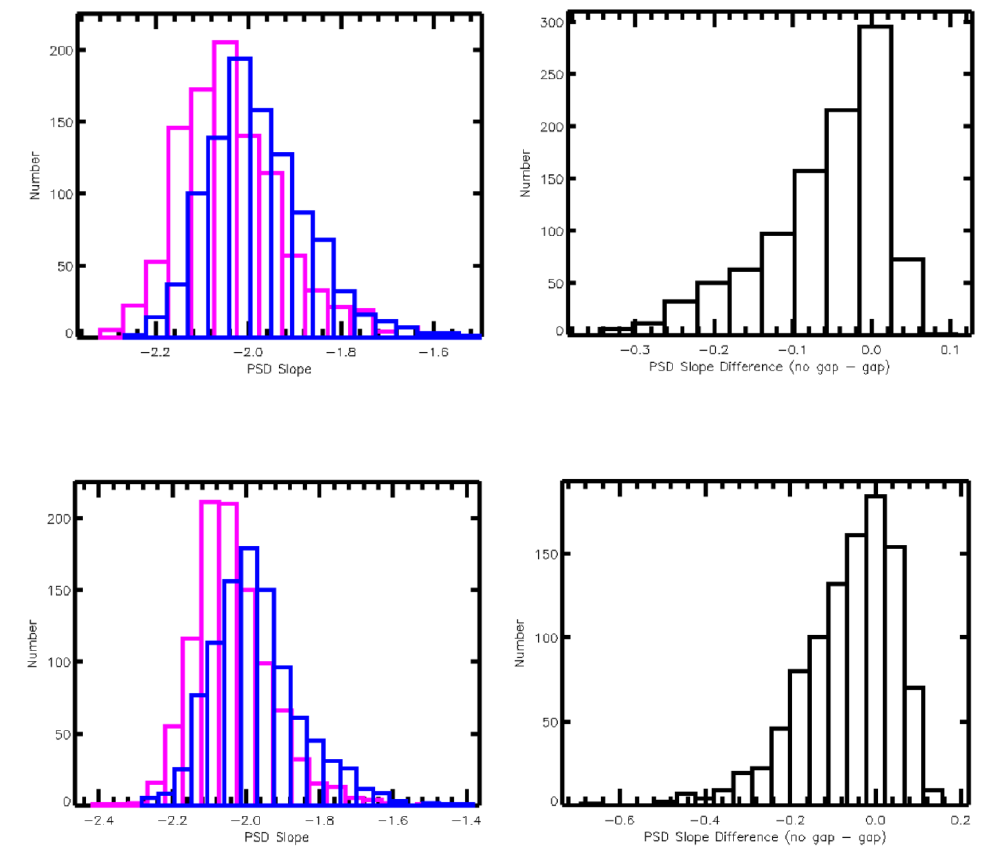}
\caption{The top left panel displays the histogram of the PSD slopes determined from the light curves with (blue) and without (magenta) a 1.2 day gap for an input slope of $-2.0$, and the right panel displays the histogram of the difference between the PSD determined from the light curves without and with a 1.2 day gap.  The bottom left  panel displays the histogram of the PSD slopes determined from the light curves with (blue) and without (magenta) a 5 day gap, and the bottom right panel displays the histogram of the difference between the PSD determined from the light curves without and with a 5 day gap.}
\end{center}
\end{figure}

\end{document}